\documentclass[12pt]{article}
\usepackage{epsfig}
\usepackage{rotating}
\usepackage{url}
\newcommand{\be}{\begin{equation}}
\newcommand{\ee}{\end{equation}}
\newcommand{\ba}{\begin{eqnarray}}
\newcommand{\ea}{\end{eqnarray}}

\begin{document}

\begin{center}
\Large{\textbf{Probability Distributions in Complex Systems}}\\~\\
\normalsize{D. Sornette} 
\end{center}
\begin{center}
\normalsize{
Department of Management, Technology and Economics,
ETH Zurich\\ 
Kreuzplatz 5, CH-8032 Zurich, Switzerland
}\\
\end{center}

\vskip 1cm

\noindent 
Article Outline\\
Glossary\\
I.       Definition of the subject and its importance\\
II.      Introduction\\
\indent	II.1 Complex systems\\
\indent		II.2 Probability distributions\\
\indent		II.3 Brief survey of probability distributions\\
\indent	\indent			II.3.1 The normal (or Gaussian) distribution\\
\indent	\indent			II.3.2 The power law distribution\\
\indent	\indent			II.3.3 The Stretched exponential distribution\\
III. The fascination with power laws\\
\indent		III.1 Statistical physics in general and the theory of critical phenomena\\
\indent		III.2 Out-of-equilibrium phase transition and self-organized critical systems (SOC)\\
\indent		III.3 Non-exhaustive list of mechanisms leading to power law distributions\\
IV. Testing for power law distributions in your data\\
V. Beyond power laws: ``Kings''\\
\indent		V.1 The standard view\\
\indent		V.2 Self-organized criticality versus criticality\\
\indent		V.3 Beyond power laws: five examples of ``kings''\\
\indent		V.4 Kings and crises in complex systems\\
VI. Future Directions\\
VII. Bibliography\\

\section*{Glossary}

\noindent
complex systems: systems with a large number of mutually interacting parts, often
open to their environment, which self-organize their internal structure and 
their dynamics with novel and sometimes surprising macroscopic
``emergent'' properties.
\vskip 0.3cm

\noindent
criticality (in physics): a state in which spontaneous fluctuations of the
order parameter occur at all scales, leading to diverging correlation length
and susceptibility of the system to external influences
\vskip 0.3cm

\noindent
power law distribution: a specific family of statistical distribution
appearing as a straight line in a log-log plot; does not possess
characteristic scales and exhibit the property of scale invariance.
\vskip 0.3cm

\noindent
self-organized criticality: when the system dynamics is attracted
spontaneously, without any obvious need for parameter tuning.
to a critical state with infinite correlation length and power law statistics.
\vskip 0.3cm

\noindent
stretched-exponential distribution: a specific family of sub-exponential
distribution interpolating smoothly between the exponential distribution
and the power law family.

\section{Definition of the subject and its importance}

This Core article  for the Encyclopedia of Complexity and System Science
(Springer Science) reviews briefly the concepts underlying complex systems
and probability distributions. The later are often taken as the first quantitative
characteristics of complex systems, allowing one to detect the possible
occurrence of regularities providing a step toward defining a classification of
the different levels of organization (the ``universality classes'').
A rapid survey covers the Gaussian law, the power law
and the stretched exponential distributions. The fascination for power laws
is then explained, starting from the statistical physics approach to critical
phenomena, out-of-equilibrium phase transitions, self-organized criticality,
and ending with a large but not exhaustive list of mechanisms leading to 
power law distributions. A check-list for testing and qualifying a power law
distribution from your data is described in 7 steps. This essay enlarges the 
description of distributions
by proposing that ``kings'', i.e., events even beyond the extrapolation 
of the power law tail, may reveal an information which is complementary 
and perhaps sometimes even more important than the power law 
distribution. We conclude a list of future directions.

\section{Introduction}

\subsection{Complex systems}

The study of out-of-equilibrium dynamics (e.g. dynamical phase transitions)
and of heterogeneous systems (e.g. spin-glasses) has progressively made popular
in physics the concept of complex systems and the importance of systemic approaches:
systems with a large number of mutually interacting parts, often
open to their environment, self-organize their internal structure and 
their dynamics with novel and sometimes surprising macroscopic
(``emergent'') properties. The complex system approach, which 
involves ``seeing'' inter-connections and relationships i.e. the whole
picture as well as the component parts, is nowadays pervasive
in modern control of engineering devices and business management.
It is also plays an increasing role in most of the scientific disciplines, including
biology (biological networks, ecology, evolution, origin of life, immunology, neurobiology,
molecular biology, etc), geology (plate-tectonics, earthquakes and volcanoes, erosion and
landscapes, climate and weather, environment, etc.), 
economics and social sciences (including cognition, distributed learning,
interacting agents, etc.). There is a growing recognition that progress in
most of these disciplines, in many of the pressing issues for our future welfare
as well as for the management of our everyday life, will need
such a systemic complex system and multidisciplinary approach.

A central property of a complex system is the possible occurrence of
coherent large-scale collective behaviors with a very rich structure, resulting from the 
repeated non-linear interactions among its constituents: the whole turns out 
to be much more than the sum of its parts.  Most complex systems around us
exhibit rare and sudden transitions, that occur over
time intervals that are short compared to the
characteristic time scales of their posterior evolution.
Such extreme events express more than anything else the underlying ``forces'' usually
hidden by almost perfect balance and thus provide the potential for a better
scientific understanding of complex systems. 
These crises have fundamental societal
impacts and range from large natural catastrophes such as earthquakes, volcanic
eruptions, hurricanes and tornadoes, landslides,avalanches, lightning strikes,
catastrophic events of environmental degradation, to
the failure of engineering structures, crashes in the stock market, social unrest
leading to large-scale strikes and upheaval, economic drawdowns on national and
global scales, regional power blackouts, traffic gridlock, diseases and
epidemics, etc. 

Given the complex dynamics of these systems, a first standard attempt to quantify
and classify the characteristics and the possible different regimes consists in
\begin{enumerate}
\item identifying discrete events,
\item measuring their sizes,
\item constructing their probability distribution.
\end{enumerate}
The interest in probability 
distributions in complex systems has the following roots.
\begin{itemize}
\item  They offer a natural metric of the relative rate of 
occurrence of small versus large events, and thus of the 
associated risks.
\item As such, they constitute essential components of risk assessment 
and prerequisites of risk management.
\item Their mathematical form can provide constraints and guidelines
to identify the underlying mechanisms at their origin and thus at
the origin of the behavior of the complex system under study.
\item This improved understanding may lead to better forecasting skills,
and even to the option (or illusion (?)) of (a certain degree of) control \cite{Satinover1,Satinover2}.
\end{itemize}

\subsection{Probability distributions}

Let us first fix some notations and vocabulary.
Consider a process $X$ whose outcome is a real number. The probability density
function $P(x)$ of $X$ (pdf also called probability distribution) is such
that the probability that $X$ is found in a small interval $\Delta x$ around
$x$ is $P(x) \Delta x$. The probability that $X$ is between $a$ and $b$ is therefore
given by the integral of $P(x)$ between $a$ and $b$:
\be
{\cal P} (a<X<b)=\int_a^b P(x) {\rm d}x ~.
\ee
The pdf $P(x)$ depends on the units used to quantity the variable $x$ and has
the dimension of the inverse of $x$, such that $P(x)\Delta x$, being a
probability i.e. a number between $0$ and $1$, is dimensionless.
In a change of variable, say $x \to y=f(x)$, the probability is invariant.
Thus, the invariant quantity is the
probability $P(x)\Delta x$ and not the pdf $P(x)$. We thus have
\be
P(x)\Delta x = P(y)\Delta y  ,
\label{changevariable}
\ee
leading to $P(y) = P(x) |{\rm d}f/{\rm d}x|^{-1}$, taking the limit of infinitesimal
intervals. By definition, $P(x) \geq 0$. It is normalized, $\int_{x_{\min}}^{x_{\max}} P(x) {\rm d}x =1$,
where $x_{\min}$ and $x_{\max}$ (often $\pm \infty$) are the smallest and largest
possible values for $x$, respectively.

The empirical estimation of the pdf $P(x)$ is usually plotted with the
horizontal axis scaled as a graded series for the measure under consideration
(the magnitude of the earthquakes, etc.)
and the vertical axis scaled for the number of outcomes or measures in each
interval of horizontal value (the earthquakes of magnitude between $1$ and $2$,
between $2$ and $3$, etc.). This implies a ``binning'' into small intervals. If
the data is sparse, the number of events in each bin becomes small and can fluctuate,
leading to a poor representation of the data.
In this case, it is useful to construct the cumulative distribution
${\cal} P_{\leq} (x)$ defined by
\be
{\cal P}_{\leq}(x)={\cal P} (X \leq x)=\int_{-\infty}^{x} P(y) {\rm d}y~,
\ee
which is much less sensitive to fluctuations. ${\cal P}_{\leq}(x)$ gives
the fraction of events with values less than or equal to $x$.
${\cal P}_{\leq}(x)$ increases monotonically with
$x$ from $0$ to $1$. Similarly, we can define the so-called
complementary cumulative (or survivor) distribution  ${\cal P}_{>}(x)=1-{\cal
P}_{\leq}(x)$.

For random variables which take only discrete values $x_1, x_2, ..., x_n$, the pdf is
made of a discrete sum of Dirac functions $(1/n)[\delta(x-x_1)+ \delta(x-x_2)+...
+ \delta(x-x_n)]$.
The corresponding cumulative distribution function (cdf) ${\cal P}_{\leq}(x)$
is a staircase. There are also more complex distributions made of continuous
cdf but which are singular with respect to the Lebesgue measure $dx$. An
example is the Cantor distribution constructed from the Cantor set
(see for instance Chapter 5 in \cite{Sornette06}). Such singular cdf is continuous but has its derivative which is zero almost everywhere: the pdf does not exist
(see e.g. \cite{Feller}).

\subsection{Brief survey of probability distributions}

Statistical physics is rich with probability distributions. The most famous is the
 Boltzmann distribution, which describes the probability that the configuration
 of the system in thermal equilibrium has a given energy.
 Its extension to out-of-equilibrium systems is the subject 
 of intense scrutiny \cite{Ruelle} ; see also Chapter 7 of \cite{Sornette06} and references therein.
Special cases include the Maxwell-Boltzmann distribution,
the Bose-Einstein distribution and the Fermi-Dirac distribution.

In the quest to characterize complex systems, two distributions have played
a leading role: the normal (or Gaussian) distribution and the power law 
distribution. 
The Gaussian distribution is the paradigm of the ``mild'' family of distributions.
In contrast, the power law distribution is the representative  of the ``wild'' family.
The contrast between ``mild'' and ``wild'' is illustrated
by the following questions.
\begin{itemize}
\item What is the probability that someone has twice your height? Essentially zero! The
height, weight and many other variables are distributed with ``mild'' pdfs with
a well-defined typical value and relatively small variations around it. The Gaussian law
is the archetype of ``mild'' distributions.

\item What is the probability that someone has twice your wealth? The answer of course
depends somewhat on your wealth but in general, there is a non-vanishing fraction of
the population twice, ten times or even one hundred times as wealthy as you are.
This was noticed at the end of the last century by Pareto, after whom the
Pareto law has been named, which describes the power law distribution 
of wealth \cite{Zaj1,Zaj2}, a typical example of ``wild'' distributions.
\end{itemize}

\subsubsection{The normal (or Gaussian) distribution}

The expression of the Gaussian probability density function of a random
variable $x$ with mean $x_0$ and standard deviation $\sigma$ reads
\be
P_G(x)  = {1 \over \sqrt{2\pi \sigma^2}} ~\exp\left(-\frac{(x-x_0)^2}{2\sigma^2}\right)  \indent
\mbox{   defined for $-\infty < x < +\infty$ }~.
 \label{loaaaagggg}
\ee

The importance of the normal distribution as a model of quantitative phenomena in the natural and social sciences can be in large part attributed to the central limit theorem. Many measurements of physical as well as social phenomena can be well approximated by the normal distribution. While the mechanisms underlying these phenomena are often unknown, the use of the normal model can be theoretically justified by assuming that many small, independent effects are additively contributing to each observation. The Gaussian distribution is also justified as the most parsimonious
choice in absence of information other than just the mean and the variance: it maximizes the
information entropy among all distributions with known mean and variance.
As a result of the central limit theorem, the normal distribution is the most widely used family of distributions in statistics and many statistical tests are based on the assumption of 
asymptotic normality of the data. In probability theory, the standard
Gaussian distribution arises as the limiting distribution of a large class of
distributions of random variables (with suitable centering and normalization)
characterized by a finite variance, which is nothing but the statement
of the central limit theorem  (see e.g. Chapter 2 in \cite{Sornette06}).

At the beginning of the twenty-first century, when power laws are often taken as the hallmark of complexity, it is interesting
to reflect on the fact that the previous giants of science in the eighteen and nineteen
centuries (Halley, Laplace, Quetelet, Maxwell and so on) considered that the
Gaussian distribution expressed a kind of universal law of nature and of society.
In particular, the Belgian astronomer Adolphe Quetelet was instrumental in 
popularizing the statistical regularities discovered by Laplace in the frame of the 
Gaussian distribution, which influenced
 the likes of John Herschel and John Stuart Mill and led
Comte to define the concept of  ``social physics.''

\subsubsection{The power law distribution}

A probability distribution function $P(x)$ exhibiting a power law tail is such that 
\be
P(x) \propto {C_{\mu} \over x^{1+\mu}} ~, \indent~~{\rm for}~x~{\rm large}~,  \label{cdakdaka}
\ee
possibly up to some large limiting cut-off. The exponent $\mu$ (also referred to
as  the ``index'')  characterizes
the nature of the tail: for $\mu<2$, one speaks of a ``heavy tail'' for which the 
variance is theoretically not defined. The scale factor $C_{\mu}$ plays a role
analogous for power laws to the role of the variance for Gaussian distributions
(see e.g. Chapter 4 in \cite{Sornette06}). In particular, it enjoys the additivity
property: the scale factor of the distribution of the sum of several independent random variables,
each with a distribution exhibiting a power law tail with the same exponent $\mu$, is equal to the
sum of the scale factors characterizing each distribution of each random variable
in the sum.

A more general form is 
\be
P(x) \propto {L(x) \over x^{1+\mu}} ~, \indent~~{\rm for}~x~{\rm large}~,  \label{cdakdaka2}
\ee
where $L(x)$ is a slowly varying function defined by lim$_{x \to \infty} L(tx)/L(x) =1$
for any finite $t$ (typically, $L(x)$ is a logarithm $\ln(x)$ or power of a logarithm such as $(\ln(x))^n$ with $n$ finite).  In mathematical language, 
a function such as (\ref{cdakdaka2})  is said to be ``regularly varying.''
This more general form means that the power law regime
is only an asymptotic statement holding as a better and better approximation
as one considers larger and larger $x$ values.

Power laws obey the symmetry of scale invariance, that is, they verify 
the following defining property that, for an arbitrary real number $\lambda$, there 
exists a real number $\gamma$ such that
\be 
P(x) = \gamma  P(\lambda x) ~,\indent~~\forall x
\label{one}
\ee
Obviously, $\gamma = \lambda^{1+\mu}$. The relation (\ref{one}) means
that the ratio of the probabilities of occurrence of two sizes $x_1$ and $x_2$
depend only on their ratio $x_1/x_2$ and not on their absolute values.
For instance, according to the Zipf law ($\mu=1$) for the distribution
of city sizes, the ratio of the number of cities with more that 1 million inhabitants
to those with more than $100'000$ persons is the same as the ratio of the number of
cities with more than $100'000$ inhabitants to those with more than $10'000$
persons, both ratios being equal to $1/10$.
The symmetry of scale invariance (\ref{one})
extends to the space of functions the concept of scale invariance
which characterizes fractal geometric objects.

It should be stressed that, when they exhibit a power law-like shape, most empirical
distributions do so only over  a finite range of event sizes, either bounded between a lower
and an upper cut-off \cite{Malcai,Biham1,Biham2,Mandel1}, or above a lower threshold, i.e., only in the tail of the observed distribution \cite{Mandel2,Ahafeder,Riste,PS03}.
Power law distributions and more generally regularly varying distributions
remain robust functional forms under a large number of operations, such as
linear combinations, products, minima, maxima, order statistics, powers,
which may also explain their ubiquity and attractiveness. 
Jessen and Mikosch \cite{Jessen} give the conditions under which transformations
of power law distributions are also regularly varying, possibly with a different 
exponent (see also \cite{Sornette06}, section 4.4 for an heuristic presentation
of similar results).

\subsubsection{The Stretched exponential distribution}

The so-called stretched exponential (SE) distributions have been found 
to be a versatile intermediate distribution interpolating between
``thin tail'' (Gaussian, exponential,...) and very ``fat tail'' distributions. In particular,
Laherr\`ere and Sornette \cite{LahSor} have found that several examples of fat-tailed
distribution in the natural and social sciences, often considered to be good
examples of power laws, could be represented as well as or even better
sometimes by a SE distribution. Malevergne et al. \cite{MalPisSor05} present systematic statistical
tests comparing the SE family with the power law distribution in the context
of financial return distributions. The SE family is defined by the following
expression for the survival distribution (also called the complementary
cumulative distribution function):
\be
\label{eq:Weibull}
                {\cal P}_{\geq u}(x)=1- \exp \left[- \left( \frac{x}{d} \right)^{c}+
\left( \frac{u}{d}\right)^{c}\right]~,\indent{\rm for}~ x \geq u~.
\ee
The constant $u$ is a lower threshold that can be changed to emphasize more the tail
of the distribution as $u$ is increased.
The structural exponent $c$ controls the ``thin'' versus ``heavy'' nature of the tail.
\begin{enumerate}
\item For $c=2$, the SE distribution (\ref{eq:Weibull}) has the same asymptotic tail
as the Gaussian distribution. 
\item For $c=1$, expression (\ref{eq:Weibull}) recovers the pure exponential distribution.
\item For $c<1$, the tail of  ${\cal P}_{u}(x)$ is fatter than an exponential, and 
corresponds to the regime of sub-exponentials (see Chapter 6 in  \cite{Sornette06}).
\item For $c \to 0$ with 
\be
c \cdot \left( \frac{u}{d}\right)^c \rightarrow \mu~, 
\label{hngr}
\ee
the SE distribution converges to the Pareto distribution with tail exponent $\mu$.
\end{enumerate}
 Indeed, we can write
\ba
\frac{c}{d^c} \cdot x^{c-1} \cdot \exp \left( - \frac{x^c-u^c}{d^c} \right) &=&
c \left( \frac{u}{d} \right)^c \cdot \frac{x^{c-1}}{u^c} \exp \left[-\left(
\frac{u}{d}\right)^c \cdot \left( \left(\frac{x}{u} \right)^c-1 \right)
\right]~,  \nonumber\\
& \simeq & \mu \cdot x^{-1} \exp \left[ - c\left( \frac{u}{d}\right)^c  \cdot
\ln \frac{x}{u} \right] , \indent	 {\rm as}~ c \rightarrow 0   \nonumber \\
& \simeq & \mu \cdot x^{-1} \exp \left[ - \mu  \cdot \ln
\frac{x}{u} \right]
~,  \nonumber \\
& \simeq& \mu \frac{u^\mu}{x^{\mu+1}}~,   \label{jgjkwlw}
\ea
which is the pdf of the Pareto power law model with tail index $\mu$. 
This implies that, as $c \to 0$, the characteristic scale $d$ of the SE model
must also go to zero with $c$ to ensure its convergence 
towards the Pareto distribution.

This shows that the Pareto model can be approximated with any desired accuracy
on an arbitrary interval $(u>0,U)$ by the (SE) model
with parameters $(c,d)$ satisfying equation (\ref{hngr}) where
the arrow is replaced by an equality. The limit $c \to 0$ provides any desired approximation to the
Pareto distribution, uniformly on any finite interval $(u,U)$.
This deep relationship between the SE and power law models allows us to
understand why it can be very
difficult to decide, on a statistical basis, which of these models fits
the data best \cite{LahSor,MalPisSor05}.
This insight can be made rigorous to develop a formal statistical test 
of the (SE) hypothesis versus the Pareto
hypothesis \cite{MalPisSor05,MalSor06}.
 
From a theoretical view point, this
class of distributions (\ref{eq:Weibull}) is motivated in part by the fact that
the large deviations of multiplicative processes are generically
distributed with stretched exponential distributions \cite{FrischSor}.
Stretched exponential distributions are also
parsimonious examples of the important subset of sub-exponentials,
that is, of the general class of distributions decaying slower than an
exponential \cite{Willekens}. This class of sub-exponentials share
several important
properties of heavy-tailed distributions \cite{Embrechts}, not shared by
exponentials or distributions decreasing faster than exponentials:
for instance,
they have ``fat tails'' in the sense of the asymptotic
probability weight of the maximum compared with the sum of large
samples \cite{Feller} (see also \cite{Sornette06}, Chapters.1 and 6).

Notwithstanding their fat-tailness, Stretched Exponential
distributions have all
their moments finite, in contrast with regularly
varying distributions for which moments of order equal to or larger
than the tail
index $\mu$ are not defined. However, they do not admit an exponential
moment, which leads to problems in the reconstruction of the distribution
from the knowledge of their moments \cite{Stuart}.
In addition, the existence of all moments is an important property
allowing for an
efficient estimation of any high-order moment, since it ensures that the
estimators are asymptotically Gaussian. In particular, for
Stretched-Exponentially distributed random variables, the variance,
skewness and kurtosis can be accurately estimated, contrarily to random
variables with regularly
varying distribution with tail index smaller than about $5$.

\section{The fascination with power laws}

Probability distribution functions with a power law dependence in 
terms of event or object sizes seem to 
be ubiquitous statistical features of
natural and social systems. It has repeatedly been argued that such an observation
relies on an underlying self-organizing mechanism, and therefore power laws
should be considered as the statistical imprints of complex systems. It is often claimed that
the observation of a power law relation in data often points to specific kinds of mechanisms
at its origin, that can often suggest a deep connection with other, seemingly unrelated systems.
In complex systems,  the appearance of power law distributions
is often thought to be the signature of hierarchy and robustness.
In the last two decades, such claims have been made for instance for earthquakes, weather and climate changes, solar flares, the fossil record, and many other systems, 
to promote the relevance of self-organized criticality as an underlying mechanism
for the organization of complex systems \cite{Bak96}. This claim is often unwarranted as there are many non-self-organizing
mechanisms producing power law distributions \cite{Sornette94,Sornette02,Sornette06,Newman05}.

Research on the origins of power law relations, and efforts to observe and validate them in the real world, is extremely active in many fields of modern science, including physics, geophysics, biology, medical sciences, computer science, linguistics, sociology, economics and more.
One can attempt to summarize briefly the present understanding as follows. 

\subsection{Statistical physics in general and the theory of critical phenomena} 

The study of critical phenomena in statistical physics
suggests that power laws emerge close to special critical or bifurcation points
separating two different phases or regimes of the system. In systems at
thermodynamic equilibrium modeled by general spin models, 
the renormalization group theory \cite{Wilson} has demonstrated the existence of universality,
so that diverse systems exhibit the same critical exponents and 
identical scaling behavior as they approach criticality, i.e., they
share the same fundamental macroscopic properties. For instance, the behavior of water and CO$_2$ at their boiling points at a certain critical pressure and that of a magnet at its Curie point fall in the same universality class because they can be characterized by the same
order parameter in the same space dimension.  In fact, almost all material phase transitions are described by a small set of universality classes.  

From this perspective, the
fascination with power laws reflects the fact that they characterize the many coexisting and delicately interacting scales at a critical point. The existence of many scales leading to complex
geometrical properties is often associated with fractals \cite{Mandel2}. While it is true that critical points  and fractals share power law relations, power law relations and
power law distributions are not the same. The later, which is the subject of this essay, describes the
probability density function or frequency of occurrence of objects or events, such as the frequency of earthquakes of a given magnitude range. In contrast, power law
relations between two variables (such as the magnetization and temperature in the case 
of the Curie point of a magnet) describe a functional abstraction belonging to or characteristic of these two variables. Both power law relations and power law distributions can result from 
the existence of a critical point. A simple example in percolation is (i) the power law dependence
of the size of the larger cluster as a function of the distance from the percolation 
threshold and (ii) the power law distribution of cluster sizes at the percolation threshold 
\cite{StaufferAha}.

\subsection{Out-of-equilibrium phase transition and self-organized critical systems (SOC)} 

In the broadest sense, SOC refers to the
spontaneous organization of a system driven from the outside into a globally
stationary state, which is characterized by self-similar
distributions of event sizes and fractal geometrical properties.  This stationary
state is dynamical in nature and is characterized by statistical fluctuations,
which are generically refered to as ``avalanches.''

The term ``self-organized criticality'' contains two parts. The word ``criticality''
refers to the state of a system at a critical point at which the correlation length
and the susceptibility become infinite in the infinite size limit as in the preceding section.
The label ``self-organized'' is often applied indiscriminately to pattern formation
among many interacting elements. The concept is that the structuration, the patterns
and large scale organization appear spontaneously.  The notion of self-organization
refers to the absence of control parameters.

In this class of mechanisms, where
the critical point is the attractor, the situation becomes more complicated as the number
of universality classes proliferates. In particular, it is not generally well-understood why
sometimes local details of the dynamics may change the macroscopic properties
completely while in other cases, the university class is robust. In fact,  the more we learn about
complex out-of-equilibrium systems, the more we realize that the concept of
universality developed for critical phenomena at equilibrium has to be enlarged to
embody a more qualitative meaning: the critical exponents defining the
universality classes are often very sensitive to many (but not all) details of the
models \cite{Gabrielov}.

Of course, one of the hailed hallmark of SOC is the existence of power law
distributions of  ``avalanches'' and of other quantities \cite{Bak96,Jensen00,Sornette06}.

\subsection{Non-exhaustive list of mechanisms leading to power law distributions}

There are many physical and/or mathematical mechanisms that generate
power law distributions and self-similar behavior. Understanding how a mechanism is
selected by the microscopic laws constitute an active field of research.
We can propose the following non-exhausive list of mechanisms that have been
found to be operating in different complex systems, and which can lead
to power law distribution of avalanches or cluster sizes. For most of these
mechanisms, we refer the reader 
to \cite{Sornette06} [Chapters 14 and 15]  and to \cite{Mitzenmacher,Newman05}
for detailed explanations and the relevant bibliography. However, some
of the mechanisms mentioned here have not been reviewed in these three references
and are thus new to the list developed in particular in  \cite{Sornette06}. We should also
stress that some of the mechanisms in this list are actually different incarnations
of the same underlying idea (for instance preferential attachment which is a re-discovery 
of the Yule process, see \cite{Simkin} for an informative historical account). 
\begin{enumerate}
\item percolation, fragmentation and other related processes, 
\item directed percolation and its universality class of so-called ``contact processes'',
\item cracking noise and avalanches resulting from the competition between frozen disorder and local interactions, as exemplified in the random field Ising model, where avalanches result from hysteretic loops \cite{Sethna},
\item random walks and their properties associated with their first passage statistics
\cite{Redner} in homogenous as well as in random landscapes,
\item flashing annihilation in Verhulst kinetics \cite{Zyg},
\item sweeping of a control parameter towards an instability \cite{Sornette94,StauSor99},
\item proportional growth by multiplicative noise with constraints (the Kesten process \cite{Kesten}
and its generalization for instance in terms of generalized Lotka-Volterra processes \cite{Solomon},
whose ancestry can be traced to Simon and Yule, 
\item competition between multiplicative noise and birth-death processes \cite{Saichevetal07},
\item growth by preferential attachment \cite{Mitzenmacher},
\item exponential deterministic growth with random times of observations (which gives the Zipf law)
\cite{Reed},
\item constrained optimization with power law constraints (HOT for highly optimized tolerant),
\item control algorithms, which employ optimal parameter estimation based on past observations, have been shown to generate broad power law distributions of fluctuations and of their corresponding corrections in the control process \cite{Cabrera,Eurich},
\item on-off intermittency as a mechanism for power law pdf of laminar phases
\cite{Platt,Heagy},

\item self-organized criticality which comes in many flavors as explained in 
Chapter 15 of \cite{Sornette06}:
	\begin{itemize}
	\item cellular automata sandpiles with and without conservation laws,
	\item systems made of coupled elements with threshold dynamics,
	\item critical de-synchronization of coupled oscillators of relaxation,
	\item nonlinear feedback of the order parameter onto the control parameter
	\item generic scale invariance,
	\item mapping onto a critical point,
	\item extremal dynamics.
	\end{itemize}
\end{enumerate} 

If there is one lesson to extract from this impressive list, it is that, when observing an
approximate linear trend in the log-log plot of some data distribution, one should
refrain from jumping to hasty conclusions on the implications of this approximate
power law behavior. Another lesson is that power laws appear to be so ubiquitous
perhaps because many roads lead to them!

\section{Testing for power law distributions in your data}

Although power law distributions are attractive for their simplicity (they are 
straight lines on log-log plots) and may be justified from theoretical reasons as 
discussed above,
demonstrating that data do indeed follow a power law distribution requires more than simply fitting. 
Indeed, several alternative functional forms can appear to follow a power law form over some extent, such as stretched exponentials and log-normal distributions. Thus, validating
that a given distribution is a power law is not easy and there is no silver bullet. 

Clauset et al. \cite{Clauset} have recently summarized some statistical techniques for making accurate parameter estimates for power-law distributions, based on maximum likelihood methods and the Kolmogorov-Smirnov statistic. They illustrate these statistical methods
on twenty-four real-world data sets from a range of different disciplines.  In some cases, 
they find that power laws are consistent with the data while in others the power law is ruled out.

Here, we offer some advices for the characterization 
of a power law distribution as a possible adequate representation of a given data set.
We emphasize good sense and practical aspects.

\begin{enumerate}
\item {\bf Survivor distribution}. First, the survival distribution should be constructed using the raw
data by ranking the values in increasing values. Then, rank versus
values gives immediately a non-normalized survival distribution. The advantage of this
construction is that it does not require binning or kernel estimation, which 
is a delicate art, as we have alluded to.

\item {\bf Probability density function}. The previous construction of the complementary cumulative (or
survivor) distribution function should be complemented with that of the density function.
Indeed, it is well-known that the cumulative 
distribution, being a ``cumulative'' integral of the density function as its name indicates, may be contaminated by disturbances
at one end of the density function, leading to rather long cross-overs that may 
hide or perturb the power law. For instance, if the 
generating density distribution is a power law truncated by an exponential,
as found for critical systems not exactly at their critical point or in the presence of finite-size effects 
\cite{Cardy}, the power law part of the cumulative distribution will be strongly 
distorted leading to a spurious estimation of the exponent $\mu$. This problem can be
in large part alleviated by constructing the pdf using binning or, even better, kernel methods
(see the very readable article \cite{Cranmer} and references therein). 
By testing and comparing the survival and the probability density distributions, one obtains
either a confirmation of the power law scaling or an understanding of the origin(s) of the deviations from the power law.

\item {\bf Structural analysis by visual inspection}.
Given that these first two steps have been performed, we recommend a preliminary
visual exploration by plotting the survival and density distributions in (i) linear-linear
coordinates, (ii) log-linear coordinates (linear abscissa and logarithmic ordinate) and (iii) log-log
coordinates  (logarithmic abscissa and logarithmic ordinate). The visual comparison
between these three plots provides a fast and intuitive view of the nature of the data.
	\begin{itemize}
	\item A power law distribution will appear as a convex curve in the linear-linear and log-linear
plots and as a straight line in the log-log plot. 
	\item A Gaussian distribution will appear
as a bell-shaped curve in the linear-linear plot, as an inverted parabola in the log-linear
plot and as strongly concave sharply falling curve in the log-log plot. 
	\item An exponential distribution will appear as a convex curve in the linear-linear
plot, as a straight line in the log-linear plot and as a concave curve in the log-log plot.
	\end{itemize}
Having in mind the shape of these three reference distributions in these three
representations provides fast and useful reference points to classify the unknown
distribution under study. For instance, if the log-linear plot shows a convex shape (upward
curvature), we can conclude that the distribution has a tail fatter than an exponential.
Then, the log-log plot will confirm if a power law is a reasonable description. If the
log-log plot shows a downward curvature (concave shape), together with the first
information that  the log-linear plot shows a convex shape,  we can conclude
that the distribution has a tail fatter than an exponential but thinner than a power law.
For example, it could be a gamma distribution ($\sim x^n~\exp[-x/x_0]$ with $n>0$) or a stretched
distribution (expression (\ref{eq:Weibull}) with $c<1$). Only more detailed quantitative
analysis will allow one to refine the diagnostic, often with not definite conclusions
(see as an illustration the detailed statistical analysis comparing the power law
to the stretched exponential distributions to describe the distribution of financial
returns \cite{MalPisSor05}).

The deviations from linearity in the log-log plot suggest
the boundaries within which the power law regime holds. We said ``suggest'' as a visual
inspection is only a first step, which can be actually misleading. While we recommend
a first visual inspection, it is only a first indication, not a proof. It is a necessary step
to convince oneself (and the reviewers and journal editors) but certainly not a sufficient
condition. It is a standard rule of thumb that a power law scaling is thought to be
meaningful if it holds over at least two to three decades on both axes and is 
bracketed by deviations on both sides whose origins can be understood (for instance,
due to insufficient sampling and/or finite-size effects).

As an illustration of the potential errors stemming from visual inspection,  we refer to the discussion of  Sornette et al., \cite{Soretal96} on the claim of Pacheco et al. \cite{Pacheco} of the existence of a break
in the Gutenberg-Richter distribution of earthquake magnitudes at $m=6.4$ 
for California. This break was
claimed to reveal the finiteness of the crust thickness according to Pacheco et al.[1992].
This claim has subsequently been
shown to be unsubstantiated, as the Gutenberg-Richter law (which is a power law when expressed
in earthquake energies or seismic moments) seems to remain valid up to magnitudes of $7.5$
in California and up to magnitude about $8-8.5$ worldwide.
This visual break at $m=6.4$ turned out to be just a statistical deviation, completely expected from the nature of power law fluctuations \cite{Main,PS03}.

\item {\bf OLS fitting}. The next step is often to perform an OLS (ordinary least-square) regression of the
data (survival distribution or kernel-reconstructed density) in the logarithm of the variables,
in order to estimate the parameters of the power law. These parameters are the exponent $\mu$,
the scale factor $C_{\mu}$ and possibly an upper threshold or other parameters controlling
the cross-over to other behaviors outside the scaling regime.
Using logarithms ensures that all the terms in  the sum of squares over the different data
points contribute approximately similarly in the OLS. Otherwise, without logarithms, 
given the large range of values spanned by a typical power law distribution, a relative error 
of say $1\%$ around a value of the order of $10^4$ would have a weight 
in the sum ten thousand times larger than the weight due to the same relative error
of $1\%$ around a value of the order of $10^2$, biasing the
estimation of the parameters towards fitting preferentially the large values. In addition, 
in logarithm units, the estimation of the exponent $\mu$ of a power law constitutes
a linear problem which is solved analytically.

\item {\bf Maximum likelihood estimation}. Using an OLS method to estimate the parameters of a power law assumes
implicitly that the distribution of the deviations from the power law (actually
the difference between the logarithm of the data and the logarithm of the power law
distribution) are normally distributed. This may not be a suitable approximation.
An estimation which removes this assumption consists in using the likelihood
method, in which the parameters of the power law are chosen so as to maximize the likelihood function.
When the data points are independent, the likelihood function is nothing
but the product $\prod_{i=1}^N P(x_i)$  over the $N$ data points
$x_1, x_2, ..., x_N$ of the power law distribution  $P(x)$. In this case, the exponent
$\mu$ which maximizes this likelihood (or equivalently and more conveniently
its logarithm called the log-likelihood) is called the Hill estimator \cite{Hill}. It reads
\be
{1 \over \mu}= {1 \over n} \sum \ln \left[ {x_j \over x_{\rm min}}\right]~, 
\ee
where $x_{\rm min}$ is the smallest value among the $n$ values used in the data set
for the estimation of $\mu$. Since power laws are often asymptotic characteristics of the tail, 
it is appropriate not use the full data set
but only the upper tail with data values above a lower threshold. Then, plotting 
$1/\mu$ or $\mu$ as a function
of the lower threshold usually provides a good sense of the existence of a power law regime:
one should expect an approximate stability of $1/\mu$ over some scaling regime. 
Note that the Hill estimator provides an unbiased estimate of $1/\mu$ while $\mu$ obtained
by inverting $1/\mu$ is slightly biased (see e.g. Chapter 6 in \cite{Sornette06}).
We refer to \cite{Drees,Resnik}  for improved versions and procedures of the Hill estimator  which deal with finite ranges and dependence. 

\item {\bf Non-parametric methods}. Methods testing for a power law behavior in a given empirical distribution which
are not parametric and sensitive provide useful complements of the above fitting and 
parametric estimation approaches. Pisarenko et al. \cite{Pisetal04} and Pisarenko and Sornette 
\cite{PisaSor06} have 
developed a new statistics, such that a power law behavior is associated with a zero value
of the statistics {\it independently of the numerical value of the exponent $\mu$} and
with a non-zero value otherwise. Plotting this statistics as a function of the lower threshold of the 
data sample allows one to detect subtle deviations from a pure power law. Lasocki \cite{Lasocki1}
and Lasocki and Papadimitriou \cite{Lasocki2} have developed another non-parametric 
approach to detect deviations from a power law,
the smoothed bootstrap test for multimodality which makes it
possible to test the complexity of the distribution without specifying any particular
probabilistic model. The method relies on testing the hypotheses that the 
number of modes or the number of bumps exhibited by the distribution function 
equal to $1$. Rejection of one of these hypotheses indicate that the distribution
has more complexity than described by a simple power law.

\end{enumerate}

Once the evidence for a power law distribution has been reasonably demonstrated,
the most difficult task remains: finding a mechanism and model which can explain
the data. Note that the term ``explain'' refers to different meanings depending on
whose expert you are speaking to. For a statistician, having been unable to reject
the power law function (\ref{cdakdaka}) given the data amounts to say that the
power law model ``explains'' the data. The emphasis of the statistician will be
on refining parametric and non-parametric procedures to test the way the power law
``fits'' or deviates from the empirical data. In contrast, a physicist or a natural scientist sees this
only as a first step, and attributes the word ``explain'' to the stage where a mechanism
in terms of a more fundamental process or first principles can derive the power law.
But even among natural scientists, there is no consensus on what is a suitable
``explanation.'' The reason stems from the different cultures and levels of study in 
different fields, well addressed in the famous paper ``More is different'' of Anderson \cite{Anderson}:
a suitable explanation for a physicist will frustrate a chemist who herself will
make unhappy a biologist. Each scientific discipline is anchored in a more
fundamental scientific level while having developed its specific concepts, which 
provide the underpinning for the next scientific level of description (think for instance
of the hierarchy:
physics $\to$ chemistry $\to$ molecular biology $\to$ cell biology $\to$ animal
biology $\to$ ethology $\to$ sociology $\to$ economics $\to$ ...).

Once a model at a given scientific description level has been proposed, 
the action of the model on inputs gives outputs which are compared with the data.
Verifying that the model, inspired by the preliminary power law evidence,
adequately fits this power law is a first step. Unfortunately, much too often,
scientists stop there and are happy to report that they have a model that
fits their empirical power law data. This is not good science. Keeping in mind the many possible mechanisms at the origin of power
law distributions reviewed above, a correct procedure
is to run the candidate model to get other predictions that
can themselves be put to test.  This validation is essential to determine the degree to
which the model is an accurate representation of the real world from the
perspective of its intended uses. Reviewing a large body of literature devoted
to the problem of validation, Sornette et al. \cite{Sornetteetal07} have proposed a synthesis in which
the validation of a
given model is formulated as an iterative construction process that mimics the
often implicit process occurring in the minds of scientists. Validation is
nothing but the progressive build-up of trust in the model, based on
putting the model to test against non-redundant novel experiments
or data, that allows one to make a decision and act decisively.
The applications of the validation program to a cellular
automaton model for earthquakes, to a multifractal random walk model for
financial time series, to an anomalous diffusion model for solar radiation
transport in the cloudy atmosphere, and to a computational fluid dynamics
code for the Richtmyer-Meshkov instability, exemplify the importance of
going beyond the simple qualification of a power law.

\section{Beyond power laws: ``Kings''}

\subsection{The standard view}

Power law distributions incarnate the notion that extreme events 
are not exceptional $9$-sigma events (to refer to the terminology
using the Gaussian bell curve and its standard deviation $\sigma$ as the
metric to quantify deviations from the mean). Instead, extreme events
should be considered as rather frequent and part of the same organization
as the other events. In this view, a great earthquake is just an earthquake that
started small ... and did not stop; it is inherently unpredictable
due to its sharing of all the properties and characteristics of smaller events 
(except for its size), so that no genuinely informative precursor can be identified
\cite{Geller}. This is the view expounded
by Bak and co-workers in their formulation of the concept of self-organized criticality
\cite{BakPac,Bak96}. In the following, we outline several 
promising directions of research that expand on these ideas.

\subsection{Self-organized criticality versus criticality}

However, there are many suggestions that this does not need to be the case.
One argument is that criticality and self-organized criticality (SOC)  can actually 
co-exist.  The hallmark of criticality is the existence of specific precursory patterns (increasing susceptibility and correlation length) in space and time. Continuing with the example
of earthquakes, the idea that a great earthquake could result from a critical phenomenon has been put forward by different groups, starting almost three decades ago 
\cite{Allegre,Keilis,SS90}. Attempts to link earthquakes and critical phenomena find support in the evidence that rupture in heterogeneous media is similar to a critical phenomenon (see Chapter 13 of \cite{Sornette06} and references therein). Also indicative is the often reported observation of increased intermediate magnitude seismicity before large events \cite{Bowman,SamSor}.
An illustration of the coexistence of criticality and of SOC is found in a simple sandpile model of earthquakes on a hierarchical fault structure \cite{Huang98}. Here, the important ingredient is to take into account both the nonlinear dynamics and the complex geometry.  From the point of view of self-organized criticality, this is surprising news: large earthquakes do not lose their identity. In the model of Huang et al. \cite{Huang98}, a large earthquake is different from a small one, a very different story than the one told by common SOC wisdom in which any precursory state of a large event is essentially identical to a precursory state of a small event and an earthquake does not know how large it will become. The difference comes from the absence of geometry in standard SOC models. Reintroducing geometry is essential. In models with hierarchical fault structures, one finds a degree of predictability of large events.

\subsection{Beyond power laws: five examples of ``kings''}

Are power laws the whole story? The following examples suggest that
some extreme events are even ``wilder'' than predicted by the
extrapolation of the power law distributions. They can be termed ``outliers''
or even better ``kings'' \cite{LahSor}. According to the definition of the Engineering 
Statistical Handbook \cite{Handbook}, ``An outlier is an observation that lies an abnormal 
distance from other values in a random sample from a population.''  Here, we follow
Laherr\`{e}re and Sornette \cite{LahSor} and use the term ``king'' to refer to events which are 
even beyond the extrapolation of the fat tail distribution of the rest of the population.

\begin{itemize}
\item {\bf Material failure and rupture processes}. There is now ample evidence
that the distribution of damage events, for instance quantified by the acoustic
emission radiated by micro-cracking in heterogeneous systems, is well-described
by a Gutenberg-Richter like power law \cite{Pollock,Omel,Fineberg,Lei}.
But consider now the energy released in the final global
event rupturing the system in pieces! This release of energy is many many times
larger than the largest ever recorded event in the power law distribution.
Material rupture exemplifies the co-existence of a power law distribution
and a catastrophic event lying beyond the power law.

\item {\bf Gutenberg-Richter law and characteristic earthquakes}. In seismo-tectonics,
the situation is muddy because of the difficulties with defining unambiguously 
the spatial domain of influence of a given fault. The researchers who have
delineated a spatial domain surrounding a clearly mapped large fault claim to
find a Gutenberg-Richter distribution up to a large magnitude region characterized
by a bump or anomalous rate of large earthquakes.  These large earthquakes
have rupture lengths comparable with the fault length \cite{Wesnou1,Wesnou2}.
If proven valid, this concept of a characteristic earthquake provides another 
example in which a ``king'' coexists with a power law distribution of smaller events.
Others have countered that this bump disappears when removing the somewhat artificial
partition of the data \cite{Kagan1,Kagan2}, so that the characteristic earthquake concept 
may be a statistical artifact. In this view, a particular fault may appear to have characteristic earthquakes, but the stress-shedding region, as a whole, behaves according to a pure scale-free power law distribution.

Several theoretical models have been offered to support the idea that, in some seismic regimes,
there is a coexistence between a power law and a large size regime (the ``king'' effect).
Gil and Sornette \cite{Gil} reported that this occurs when the characteristic rate for local
stress relaxations is fast compared with the diffusion of stress within the system.
The interplay between dynamical effects and heterogeneity has also been shown to 
change the Gutenberg-Richter behavior to a distribution of small events combined 
with characteristic system size events \cite{Fisher,Benzion,Hillers,Zoller}. On the empirical side, progress should be made in testing
the characteristic earthquake hypothesis by using the prediction of the models
to identify independently of seismicity those seismic regions in which the king effect 
is expected. This remains to be done [Ben-Zion, private communication, 2007].

\item {\bf Extreme king events in the pdf of turbulent velocity fluctuations}.
The evidence for kings does not require 
and is not even synonymous in general with the existence of
a break or of a bump in the distribution of event sizes. This point 
is well-illustrated in the shell models of turbulence, that are believed to capture the
essential ingredient of these flows, while being amenable
to analysis. Such ``shell'' models replace the three-dimensional
spatial domain by a series of uniform onion-like spherical layers 
with radii increasing as a geometrical series $1, 2, 4, 8, ..., 2^n$
and communicating with each other mostly with nearest neighbors.
The quantity of interest
is the distribution of velocity variations between two instants at
the same position or between two points simultaneously. 
L'vov et al. \cite{Lvov} have shown that they could collapse the pdf's
of velocity fluctuations for different scales only for the small
velocity fluctuations, while no scaling held for large velocity 
fluctuations. The conclusion is that the distributions
of velocity increments seem to be composed of two regions, a region of so-called
``normal scaling'' and a domain of extreme events.
They could also show that these extreme 
fluctuations of the fluid velocity correspond to intensive
peaks propagating coherently (like solitons) over several shell layers with a
characteristic bell-like shape, approximately independent of 
their amplitude and duration (up to a rescaling of their size
and duration).  One could summarize these findings by 
saying that ``characteristic'' velocity pulses decorate an otherwise
scaling probability distribution function.

\item {\bf Outliers and kings in the distribution of financial drawdowns}.
In a series of papers, Johansen and Sornette \cite{Johansen1,Johansen2,Johansen3} have shown that 
the distribution of drawdowns in financial markets exhibits this coexistence
of a fat tail with a characteristic regime with ``kings'' (called ``outliers'' in the papers).
The analysis encompasses exchange markets (US dollar against the Deutsch Mark and against 
the Yen), the major world stock markets, the U.S. and Japanese
bond markets and commodity markets. Here, drawdowns
are defined as a continuous decrease 
in the value of the price at the close of each successive trading day. 
The results are found robust with using
 ``coarse-grained drawdowns,'' which allows for a certain degree of fuzziness
in the definition of cumulative losses. Interestingly, the pdf of returns
at a fixed time scale, usually the daily returns, do not exhibit any anomalous
king behavior in the tail: the pdf of financial returns at fixed time scales seem
to be adequately described by power law tails \cite{Gopi}.
The interpretation proposed by Johansen and Sornette
is that these drawdown kings are associated with crashes, which occur
due to a global instability of the market which amplifies the normal behavior
via strong positive feedback mechanisms \cite{dscrash}.

\item {\bf Paris as the king in the Zipf distribution of French city sizes}.
Since Zipf \cite{Zipf}, it is well-documented that the distribution of city sizes
(measured by the number of inhabitants) is, in many countries, a power law
with an exponent $\mu$ close to $1$. France is not an exception as it
exhibits a nice power law distribution of city sizes... except for
Paris which is completely out of range, a genuine king with a size
several times larger than expected from the distribution of the rest
of the population of cities \cite{LahSor}. This king effect
reveals a particular historical organization of France, whose roots
are difficult to unravel. Nevertheless, we think that this king effect incarnated
by Paris is a significant signal to explain in order to understand the competition
between cities in Europe.
\end{itemize}

\subsection{Kings and crises in complex systems}

We propose that these kings may reveal an information which is complementary
and perhaps sometimes even more important than the power law pdf. 

Indeed, it is essential to realize that
the long-term behavior of these complex systems is
often controlled in large part by these rare catastrophic events: the universe
was probably born during an extreme explosion (the ``big-bang''); the nucleosynthesis of
all important atomic elements constituting our matter results from the colossal
explosion of supernovae; the largest
earthquake in California repeating about once every two centuries accounts for
a significant fraction of the total tectonic deformation; landscapes are more
shaped by the ``millenium'' flood that moves large boulders  than by the action
of all other eroding agents; the largest volcanic eruptions lead to major 
topographic changes as well as severe climatic disruptions; evolution is characterized by phases
of quasi-statis interrupted by episodic bursts of activity and destruction; 
financial crashes can destroy in an instant trillions
of dollars; political crises and revolutions shape the long-term geopolitical 
landscape; even our personal life is shaped on the long run by a few key
``decisions/happenances''. 

The outstanding scientific question is thus how such large-scale patterns
of catastrophic nature might evolve from a series of interactions on the smallest
and increasingly larger scales. In complex systems, it has been found that the
organization of spatial and temporal correlations do not stem, in general, from a
nucleation phase diffusing across the system. It results rather from a
progressive and more global cooperative process occurring over the whole system
by repetitive interactions. An instance would be the many occurrences of
simultaneous scientific and technical discoveries signaling the global nature of
the maturing process. 

Standard models and simulations of scenarii of 
extreme events are subject to numerous sources of error, 
each of which may have a negative impact on the validity of the predictions \cite{Karplus}.
Some of the uncertainties are under control in the modelling
process; they usually involve trade-offs between a more faithful description
and manageable calculations. Other sources of errors are beyond control as they are
inherent in the modeling methodology of the specific disciplines. The two known 
strategies for modelling are both limited in this respect\,: analytical theoretical
predictions are out of reach for most complex problems, while brute force numerical
resolution of the equations (when they are known) or of scenarii is reliable in the
``center of the distribution'', i.e. in the regime far from the extremes where
good statistics can be accumulated. Crises are extreme events that occur rarely,
albeit with extraordinary impact, and are thus completely under-sampled and
poorly constrained. Even the introduction of teraflop (or even pentaflops in the near futur)
supercomputers does not change qualitatively this fundamental limitation.

Recent developments suggest that non-traditional approaches, based on the concepts and
methods of statistical and nonlinear physics could provide a middle way to
direct the numerical resolution of more
realistic models and the identification of relevant signatures of impending
catastrophes. Enriching the concept of self-organizing criticality, 
the predictability of crises would then rely on the fact that they 
are fundamentally outliers, e.g. large earthquakes
are not scaled-up versions of small earthquakes but the result of
specific collective amplifying mechanisms. To address this challenge,
the available theoretical tools comprise in particular bifurcation and
catastrophe theories, dynamical critical phenomena and the renormalization group,
nonlinear dynamical systems, and the theory
of partially (spontaneously or not) broken symmetries. Some encouraging results 
have been gathered on concrete problems, such as the prediction of the
failure of complex engineering structures, the detection of precursors of
stock market crashes and of human parturition, with exciting potential for earthquakes.
At the beginning of the third millenium,
it is tempting to extrapolate and forecast that a larger multidisciplinary
integration of the physical sciences together with artifical intelligence and soft-computational
techniques, fed by analogies and fertilization accross the natural sciences,
 will provide a better understanding of the limits of
predictability of catastrophes 
and adequate measures of risks for a more harmonious and sustainable futur of our complex
world.

\section{Future Directions}

Our exposition has mainly focused on the concept of distributions of event sizes,
as a first approach to characterize the organization of complex systems. But, probability
distribution functions are just one-point statistics and thus provide only
an incomplete picture of the organization of complex systems. 
This opens the road to several better measures of the organization of 
complex systems.

\begin{itemize}
\item Statistical estimations of probability distribution functions is a delicate art.
An active research field in mathematical statistics which is insufficiently used
by practitioners of other sciences is the domain of ``robust estimation.''
Robust estimation
techniques are methods which are insensitive to small departures
from the idealized assumptions which have been used to optimize the
algorithm. Such techniques include M-estimates (which follow from
maximum likelihood considerations), L-estimates (which are linear
combinations of order statistics), and R-estimates (based on statistical
rank tests) \cite{Huber,Vaartetal,Wilcox}.

\item Ideally,  one would like to measure the full multivariate distribution of events, 
which can be in full generality decomposed into the set of marginal distributions
discussed above and of the copula of the system. 
A copula embodies completely the entire dependence structure
of the system \cite{Joe,Nelsen}. Copulas have recently become fashionable
in financial mathematics and in financial engineering \cite{MalSor06}.
Their use in other fields in the natural sciences is embryonic but can be 
expected to blossom.

\item When analyzing a complex system, a common trap is to assume
without critical thinking and testing that the statistics is stationary, so that
monovariate (marginal) and multivariate distribution functions are sufficient
to fully characterize the system.  It is indeed a common
experience that the dependence estimated and predicted by standard
models change dramatically at certain times. In other words, the statistical
properties are conditional on specific regimes. The existence of regime-dependent
statistical properties has been discussed in particular in climate science,
in medical sciences and in financial economics. In the later, a quite
common observation is that investment strategies, which have some
moderate beta (coefficient of regression to the market) for normal
times, can see their beta jumps to a much larger value (close to $1$ or
larger depending on the leverage of the investment) at certain times
when the market collectively dives. Said differently, investments which
are thought to be hedged against negative global market trends may
actually lose as much or more than the global market, at certain times
when a large majority of stocks plunge simultaneously. In other words,
the dependence structure and the resulting distributions at different
time scales may change in certain regimes.

The general problem of the application of mathematical statistics to
non-stationary data (including non-stationary time series) is very
important, but alas, not much can be done. There are only a few
approaches which may be used and only in specific conditions, which 
we briefly mention. 
\begin{enumerate}
\item Use of algorithms and methods which are robust
with respect to possible non-stationarity in data, such as normalization
procedures or the use of quantile samples instead of initial samples.
\item Model non-stationarity by some low-frequency random processes,
such as, {\it e.g.}, a narrow-band random process $X(t) =A(t)
\cos(\omega t +\phi(t))$ where $\omega \ll 1$ and $A(t)$ and phase
$\phi(t)$ are slowly varying amplitude and phase. In this case, the
Hilbert transform can be very useful to characterize $\phi(t)$
non-parametrically.
\item  The estimation of the parameters of a low-frequency process based
on a ``short'' realization is often hopeless. In this case, the only
quantity which can be evaluated is the uncertainty (or scatter) of the
results due to the non-stationarity.
\item 
Regime Switching popularized by Hamilton \cite{Hamilton89} for
autoregressive time series models is a special case of non-stationary,
which can be handled with specific methods. 
\end{enumerate}

\item We already discussed the problem of ``kings.'' One key issue
that needs more scrutiny is that these outliers are often identified
only with metrics adapted to take into account transient increases of
the time dependence in the time series of returns of individual
financial assets \cite{Johansen2} (see also Chap.~3 of \cite{dscrash}). These
outliers seem to belong to a statistical population which is different
from the bulk of the distribution and require some additional
amplification mechanisms active only at special times. 
The presence of such outliers both in marginal distributions and in
concomitant events, together with the strong impact of crises and of
crashes in complex systems, suggests the need for novel measures of dependence between
different definitions of events and other time-varying metrics across different variables. This
program is part of the more general need for a joint multi-time-scale
and multi-variate approach to the statistics of complex systems. 

\item The presence of outliers 
poses the problem of exogeneity versus endogeneity.
An event identified as anomalous could perhaps be cataloged as resulting
from exogenous influences. 
The concept of exogeneity is
fundamental in statistical estimation  \cite{EngleExogeneity,Bookexogeneity}. Here, we refer
to the question of exogeneity versus endogeneity in the broader context
of self-organized criticality, inspired in particular
from the physical and natural sciences. As we already discussed,
according to self-organized
criticality, extreme events are seen to be endogenous, in contrast with
previous prevailing views (see for instance the discussion in
\cite{BakPac,Sorcat}). But, how can one assert with 100\% confidence
that a given extreme event is really due to an endogenous
self-organization of the system, rather than to the response to an
external shock? Most natural and social systems are indeed continuously
subjected to external stimulations, noises, shocks, solicitations,
forcing, which can widely vary in amplitude. It is thus not clear a
priori if a given large event is due to a strong exogenous shock, to the
internal dynamics of the system, or maybe to a combination of both.
Addressing this question is fundamental for understanding the relative
importance of self-organization versus external forcing in complex
systems and underpins much of the problem of dependence between
variables. The concepts of endogeneity and exogeneity
have many applications in the natural and social sciences (see
\cite{endoexoreview} for a review) and we expect this view point to develop
into a general strategy of investigation.

\end{itemize}

{}


\vskip 1cm


\begin{thebibliography}{}



\bibitem{Satinover1} Satinover~JB and Sornette~D (2007)
``Illusion of Control'' in Minority and Parrondo Games,
European Journal of Physics B
(\url{http://arxiv.org/abs/0704.1120})

\bibitem{Satinover2} Satinover~JB and Sornette~D (2007)
Illusion of Control in a Brownian Game, Physica A
(\url{http://arxiv.org/abs/physics/0703048})

\bibitem{Sornette06} Sornette~D (2006) Critical Phenomena in Natural Sciences,
Chaos, Fractals, Self-organization and Disorder: Concepts and Tools.
2nd ed., Springer Series in Synergetics, Heidelberg.

\bibitem{Feller} Feller~W. (1971) An Introduction to Probability Theory and
its Applications. vol. II, John Wiley and sons,~New York.

\bibitem{Ruelle} Ruelle~D (2004) Conversations on Nonequilibrium Physics With an Extraterrestrial.
Physics Today, May, 48-53.

\bibitem{Zaj1} Zajdenweber~D (1976) Hasard et Pr\'evision. Economica, Paris.

\bibitem{Zaj2} Zajdenweber~D. (1997) Scale invariance in Economics and Finance.
in Dubrulle~B,~Graner~F. and Sornette~D. eds.
Scale Invariance and Beyond. EDP Sciences and Springer,~Berlin, pp.185--194.

\bibitem{Malcai} Malcai~O, Lidar~DA, Biham~O and Avnir~D (1997)
Scaling range and cutoffs in empirical fractals, Phys. Rev. E 56,~2817--2828.

\bibitem{Biham1} Biham~O, Malcai~O, Lidar~DA and Avnir~D (1998)
Is nature fractal? - Response, Science 279,~785--786.

\bibitem{Biham2} Biham~O, Malcai~O, Lidar~DA and Avnir~D (1998)
Fractality in nature - Response, Science  279,~1615--1616.

\bibitem{Mandel1} Mandelbrot,~BB (1998) Is nature fractal? Science 279,~783--784.

\bibitem{Mandel2} Mandelbrot~BB (1982)
The fractal Geometry of Nature. W.H. Freeman,~San Francisco.

\bibitem{Ahafeder} Aharony~A and Feder~J, eds. (1989) Fractals in Physics.
Physica D 38, nos. 1--3 (North Holland, Amsterdam).
 
\bibitem{Riste} Riste~T and Sherrington~D, eds. (1991)
Spontaneous Formation of Space-Time
Structures and Criticality.  Proc. NATO ASI,~Geilo,~Norway, (Kluwer,~Dordrecht.

\bibitem{PS03} Pisarenko~ VF and Sornette~D (2003)
Characterization of the frequency of extreme events by the Generalized
Pareto Distribution, Pure and Applied Geophysics 160 (12), 2343-2364.

\bibitem{Jessen} Jessen~AH and Mikosch~T (2006) Regularly varying functions,
Publications de l'Institut Mathematique, Nouvelle serie 79 (93), 1-23
 (\url{http://www.math.ku.dk/~mikosch/Preprint/Anders/jessen_mikosch.pdf})

\bibitem{LahSor} Laherr\`{e}re~J. and Sornette~D (1999)
Stretched exponential distributions in nature and
economy: Fat tails with characteristic scales, European Physical
Journal B 2, 525-539.

\bibitem{MalPisSor05} Malevergne~Y, Pisarenko~VF and Sornette~D (2005)
Empirical Distributions of Log-Returns: between the Stretched
Exponential and the Power Law?
Quantitative Finance 5 (4), 379-401.

\bibitem{MalSor06} Malevergne~Y and Sornette~D (2006)
Extreme Financial Risks
(From Dependence to Risk Management), Springer, Heidelberg.

\bibitem{FrischSor} Frisch~ U and Sornette~D (1997)
Extreme deviations and applications, Journal de Physique I, France 7,1155-1171.

\bibitem{Willekens} Willekens~E (1988) The structure of the class
of subexponential distributions, Probability Theory \& Related Fields 77, 567-581.

\bibitem{Embrechts} Embrechts~P, Kl\"uppelberg~CP and Mikosh~T (1997)
Modelling Extremal Events, Springer-Verlag, Berlin.

\bibitem{Stuart} Stuart~A and Ord~K (1994) Kendall's
advances theory of statistics, John Wiley and Sons, New York.

\bibitem{Bak96} Bak~P (1996) How Nature Works:
the Science of Self-organized Criticality. Copernicus,~New York.

\bibitem{Sornette94} Sornette~D (1994)
Sweeping of an instability : an alternative to self-organized criticality to get power laws 
without parameter tuning, J.Phys.I France 4, 209-221.

\bibitem{Sornette02} Sornette~D (2002)
Mechanism for Power laws without Self-Organization, 
Int. J. Mod. Phys. C 13 (2), 133-136.

\bibitem{Newman05} Newman~MEJ (2005) Power laws, Pareto distributions and Zipf's law,
Contemporary Physics 46, 323-351.

\bibitem{Wilson} Wilson~KG (1979) Problems in physics with many scales of
length, Scientific American 241, August,~158-179.

\bibitem{StaufferAha} Stauffer~D and Aharony~A (1994) Introduction to Percolation
Theory, 2nd ed., Taylor \& Francis,~London; Bristol,~PA.

\bibitem{Gabrielov} Gabrielov~A,~Newman~WI and Knopoff~L (1994)
Lattice models of Fracture: Sensitivity to the Local Dynamics, 
 Phys. Rev. E  50, 188-197.

\bibitem{Jensen00}  Jensen~HJ (2000) Self-Organized Criticality: Emergent Complex Behavior in Physical and Biological Systems, Cambridge Lecture Notes in Physics, Cambridge University Press. 

\bibitem{Mitzenmacher} Mitzenmacher~M (2004) A brief history of generative models
for power law and lognormal distributions, Internet Mathematics 1, 226-251.

\bibitem{Simkin} Simkin~MV and Roychowdhury~VP (2006)
Re-inventing Willis, preprint (\url{http://arXiv.org/abs/physics/0601192})

\bibitem{Sethna} Sethna~JP (2006) Crackling Noise and Avalanches: Scaling, Critical Phenomena, and the Renormalization Group, Lecture notes for Les Houches summer school on Complex Systems,
 summer 2006.
 
\bibitem{Redner} Redner~S (2001) A Guide to First-Passage Processes,
Cambridge University Press.

\bibitem{Zyg} Zygad\l{}o~R (2006)
 Flashing annihilation term of a logistic kinetic as a mechanism leading
  to Pareto distributions, preprint \url{http://arXiv.org/abs/cond-mat/0602491}

\bibitem{StauSor99} Stauffer~D and Sornette~D (1999)
Self-Organized Percolation Model for Stock Market Fluctuations, Physica A 271, N3-4, 496-506.

\bibitem{Kesten} Kesten~H (1973) Random difference equations and renewal theory for
products of random matrices, Acta Math 131, 207-248.

\bibitem{Solomon} Solomon~S and Richmond~P (2002)
Stable power laws in variable economies; Lotka-Volterra implies Pareto-Zipf,
Eur. Phys. J. B 27, 257-261.

\bibitem{Saichevetal07} Saichev~A, Malevergne~A and Sornette~D (2007) Zipf law from proportional
growth with birth-death processes, working paper.

\bibitem{Reed} Reed~WJ and Hughes~BD (2002) From Gene Families and Genera to
Incomes and Internet File Sizes: Why Power Laws are so Common in Nature,
Physical Review E, 66, 067103.

\bibitem{Cabrera} Cabrera~JL and Milton~JG (2004)
Human stick balancing: Tuning L\'evy flights to improve balance control,
Chaos 14 (3), 691-698.

\bibitem{Eurich} Eurich~CW and Pawelzik~K (2005) Optimal Control Yields Power Law Behavior,
Int.Conference on Artificial Neural Networks 2, 365-370.

\bibitem{Platt} Platt~N, Spiegel~EA and Tresser~C (1993)
 On-off intermittency: A mechanism for bursting 
Phys. Rev. Lett. 70, 279-282.

\bibitem{Heagy} Heagy~JF, Platt~N and Hammel~SM (1994) Characterization of on-off intermittency,
Phys. Rev. E 49, 1140-1150.

\bibitem{Clauset} Clauset~A, Shalizi~CR and Newman~MEJ (2007)
Power-law distributions in empirical data,
\url{http://arxiv.org/abs/0706.1062}

\bibitem{Cardy} Cardy~JL (1988) Finite-Size Scaling, North Holland, Amsterdam.

\bibitem{Cranmer} Cranmer~K (2001) Kernel estimation in high-energy physics,
Computer Physics Communications  136  (3), 198-207.

\bibitem{Soretal96} Sornette~D, Knopoff~L, Kagan~YY and Vanneste~C (1996)
Rank-ordering statistics of extreme events: application to the distribution of
large earthquakes, J.Geophys.Res. 101, 13883-13893.

\bibitem{Pacheco} Pacheco~JF, Scholz~C and Sykes~L (1992)
Changes in frequency-size relationship from small to large earthquakes, 
Nature 355, 71-73.

\bibitem{Main} Main~I (2000) Apparent Breaks in Scaling in the Earthquake Cumulative Frequency-magnitude Distribution: Fact or Artifact? Bull. Seismol. Soc. Am. 90, 86-97.

\bibitem{Hill} Hill~BM (1975) A simple general approach to
inference about the tail of a distribution,  Annals of statistics 3, 1163-1174.

\bibitem{Drees} Drees~H, de Haan~L and Resnick~SI (2000)
How to Make a Hill Plot, The Annals of Statistics 28 (1),  254-274.

\bibitem{Resnik} Resnik~SI (1997) Discussion of the Danish Data on Large Fire Insurance Losses,
Astin Bulletin 27 (1), 139-152.

\bibitem{Pisetal04} Pisarenko~VF, Sornette~D and Rodkin~M (2004)
A new approach to characterize deviations
in the seismic energy distribution from 
the Gutenberg-Richter law, Computational Seismology 35, 138-159.

\bibitem{PisaSor06} Pisarenko~VF, Sornette~D (2006)
New statistic for financial return distributions: power law or
exponential? Physica A 366, 387-400.

\bibitem{Lasocki1} Lasocki~S (2001) Quantitative evidences of complexity of magnitude
distribution in mining-induced seismicity: Implications for hazard evaluation,
in 5th International Symposium on Rockbursts and Seismicity in
Mines, ``Dynamic Rock Mass Response to Mining,'' Symp. Ser., vol.
S27, edited by G. van Aswegen, R. J. Durrheim, and W. D. Ortlepp.
pp. 543Ð 550, S. Afr. Inst. of Min. and Metall., Johannesburg.

\bibitem{Lasocki2} Lasocki~S and Papadimitriou~EE (2006) 
Magnitude distribution complexity revealed in seismicity from Greece,
J. Geophys. Res., 111, B11309, doi:10.1029/2005JB003794.

\bibitem{Anderson} Anderson PW (1972) More is different
(Broken symmetry and the nature of the hierarchical
structure of science), Science 177, 393-396.

\bibitem{Sornetteetal07} Sornette~D, Davis~AB, Ide~K, Vixie~KR, Pisarenko~VF and Kamm~JR (2007)
Algorithm for Model Validation: Theory and Applications,
Proc. Nat. Acad. Sci. USA 104 (16), 6562-6567.

\bibitem{Geller} Geller~RG, Jackson~DD, Kagan~YY and Mulargia~F (1997) 
Earthquakes cannot be predicted, Science 275 (5306), 1616-1617.

\bibitem{BakPac}  Bak, P.  and M. Paczuski (1995) Complexity, contingency and
criticality. {\it Proceedings of the National Academy of Science USA} 
{\bf 92},~6689--6696.

\bibitem{Allegre} All\`egre~CJ, Le Mouel~JL and Provost~A (1982)
Scaling rules in rock fracture and possible implications for earthquake predictions,
Nature 297, 47-49.

\bibitem{Keilis} Keilis-Borok~V (1990) The lithosphere of the Earth as a large nonlinear system, Geophys. Monogr. Ser. 60, 81-84.

\bibitem{SS90} Sornette~A and Sornette~D (1990)
Earthquake rupture as a critical point: Consequences for telluric precursors, 
Tectonophysics 179, 327-334.

\bibitem{Bowman}  Bowman~DD, Ouillon~G, Sammis~CG, Sornette~A and Sornette~D (1996)
An observational test of the critical earthquake concept, J. Geophys. Res. 103, 24359-24372.

\bibitem{SamSor} Sammis~SG and Sornette~D (2002)
Positive Feedback, Memory and the Predictability of Earthquakes,
Proceedings of the National Academy of Sciences USA 99 (SUPP1), 2501-2508.

\bibitem{Huang98} Huang~Y,  Saleur~H, Sammis~CG and Sornette~D (1998)
Precursors, aftershocks, criticality and self-organized criticality, Europhys. Lett. 41, 43-48.

\bibitem{Handbook} Engineering Statistical Handbook, National Institute 
of Standards and Technology (2007)
See \url{http://www.itl.nist.gov/div898/handbook/prc/section1/prc16.htm}

\bibitem{Pollock} Pollock~AA(1989) Acoustic Emission Inspection, Metal Handbook, Ninth edition, Vol 17, Nondestructive
Evaluation and Quality Control (ASM International), 278-294.

\bibitem{Omel} Omeltchenko~A, Yu~J, Kalia~RK and Vashishta~P (1997) Crack Front Propagation and
Fracture in a Graphite Sheet: a Molecular-dynamics Study on Parallel Computers, Phys. Rev. Lett. 78,
2148-2151.

\bibitem{Fineberg} Fineberg~J and Marder~M (1999) Instability in Dynamic Fracture, Physics Reports 313, 2-108.

\bibitem{Lei} Lei~X, Kusunose~K, Rao~MVMS, Nishizawa~O and Sato~T (2000)  Quasi-static Fault Growth
and Cracking in Homogeneous Brittle Rock Under Triaxial Compression Using Acoustic Emission
Monitoring, J. Geophys. Res. 105, 6127-6139.

\bibitem{Wesnou1} Wesnousky, SG (1994) he Gutenberg-Richter or characteristic earthquake distribution, which is it?
Bulletin of the Seismological Society of America 84 (6), 1940-1959.

\bibitem{Wesnou2} Wesnousky, SG (1996) Reply to Yan Kagan's comment on ``The Gutenberg-Richter or characteristic earthquake distribution, which is it?''
Bulletin of the Seismological Society of America 86 (1A), 286-291.

\bibitem{Kagan1} Kagan~YY (1993) Statistics of characteristic earthquakes,
Bulletin of the Seismological Society of America  83 (1), 7-24.

\bibitem{Kagan2} Kagan~YY (1996) Comment on ``The Gutenberg-Richter or characteristic earthquake distribution, which is it?'' by S. G. Wesnousky, Bull. Seismol. Soc. Am., 86, 274-285.

\bibitem{Gil} Gil~G and Sornette~D (1996)
Landau-Ginzburg theory of self-organized criticality, Phys. Rev.Lett. 76, 3991-3994.

\bibitem{Fisher} Fisher~DS, Dahmen~K, Ramanathan~S and Ben-Zion~Y (1997) Statistics of Earthquakes in Simple Models of Heterogeneous Faults, Phys. Rev. Lett., 78, 4885-4888.

\bibitem{Benzion} Ben-Zion~Y, Eneva~M and Liu~Y (2003) Large Earthquake Cycles And Intermittent Criticality On Heterogeneous Faults Due To Evolving Stress And Seismicity, J. Geophys. Res.108 (B6), 2307, doi:10.1029/2002JB002121.

\bibitem{Hillers} Hillers~G,Mai~PM, Ben-Zion~Y and Ampuero~ J-P (2007)
Statistical Properties of Seismicity Along Fault Zones at Different Evolutionary Stages,
Geophys. J. Int. 169, 515-533.

\bibitem{Zoller} Z\"oller~G, Ben-Zion~Y and Holschneider~M (2007) Estimating recurrence times and seismic hazard of large earthquakes on an individual fault, Geophys. J. Int., in press, doi: 10.1111/j.1365-246X.2007.03480.x (\url{http://earth.usc.edu/~ybz/})

\bibitem{Lvov} L'vov~VS, Pomyalov~A and Procaccia~I (2001) 
Outliers, Extreme Events and Multiscaling,
Physical Review E 6305 (5), 6118, U158-U166.

\bibitem{Johansen1} Johansen~A and Sornette~D (1998)
Stock market crashes are outliers, European Physical Journal B 1, 141-143.

\bibitem{Johansen2} Johansen~A and Sornette~D (2001)
Large Stock Market Price Drawdowns Are Outliers, 
Journal of Risk 4 (2), 69-110, Winter 2001/02
(\url{http://arXiv.org/abs/cond-mat/0010050})

\bibitem{Johansen3} Johansen~A and Sornette~D (2007)
Shocks, Crash and Bubbles in Financial Markets,
in press in Brussels Economic Review on Non-linear Financial Analysis
149-2/Summer 2007 (\url{http://arXiv.org/abs/cond-mat/0210509})
 
\bibitem{Gopi} Gopikrishnan~P, Meyer~M, Amaral~LAN and Stanley~HE (1998)
Inverse cubic law for the distribution of stock price variations, Eur. Phys. J. B 3, 139-140.

\bibitem{dscrash} Sornette, D. (2003)
{\it Why Stock Markets Crash, Critical Events in Complex Financial Systems}.
Princeton University Press, Princeton, NJ.

\bibitem{Zipf} Zipf~GK (1949) Human behaviour and the principle of least-effort, Addison-Wesley, Cambridge MA.

\bibitem{Karplus} Karplus~WJ (1992) The Heavens are Falling: The Scientific Prediction of
Catastrophes in Our Time, Plenum.

\bibitem{Huber} Hubert, P. J. (2003) {\it Robust Statistics}. Wiley-Intersecience. 

\bibitem{Vaartetal} van der Vaart, A. W., R. Gill, B. D. Ripley, S. Ross, B. Silverman and M. Stein (2000) {\it Asymptotic Statistics}. Cambridge University Press.

\bibitem{Wilcox} Wilcox,R.R. (2004) {\it Introduction to Robust Estimation and Hypothesis Testing}, 2nd Edition. Academic Press

\bibitem{Joe} Joe, H. (1997)
{\it Multivariate models and dependence concepts}. Chapman \& Hall, London.

\bibitem{Nelsen} Nelsen, R.B. (1998)
{\it An Introduction to Copulas}, Lectures Notes in statistic {\bf 139}.
Springer Verlag, New York.

\bibitem{Hamilton89} Hamilton, J.D. (1989) A New Approach to the
Economic Analysis of
Non-stationary Time Series and the Business Cycle. {\it
Econometrica} {\bf 57},~357--384.

\bibitem{EngleExogeneity} Engle, R.F., D.F. Hendry and J.-F. Richard (1983)
Exogeneity. {\it Econometrica} {\bf 51},~277--304.

\bibitem{Bookexogeneity} Ericsson, N. and J.S. Irons (1994)
{\it Testing exogeneity}, Advanced Texts in Econometrics, Oxford 
University Press, Oxford.

\bibitem{Sorcat} Sornette, D. (2002)
Predictability of catastrophic events: material rupture,
earthquakes, turbulence, financial crashes and human birth,
{\it Proceedings of the National Academy of  Science USA} {\bf 99} 
SUPP1,~2522--2529.

\bibitem{endoexoreview} D. Sornette,
Endogenous versus exogenous origins of crises,
in the monograph entitled ``Extreme Events in Nature and Society,'' 
Series:  The Frontiers Collection 
S. Albeverio, V. Jentsch and H. Kantz, eds.  (Springer, Heidelberg, 2005)
(e-print at \url{http://arxiv.org/abs/physics/0412026}).

\end{thebibliography}
\end{document}